\documentclass[a4paper,prd,preprintnumbers,twocolumn,superscriptaddress,nofootinbib,amsmath,amssymb]{revtex4-2}

\usepackage{color,hyperref}
\usepackage{times}
\usepackage{mathrsfs}
\hypersetup{colorlinks=true,linkcolor=blue,citecolor=red,filecolor=magenta,urlcolor=blue}

\usepackage{enumitem}
\usepackage{graphicx,subfigure}
\usepackage{dcolumn}
\usepackage{bm}
\usepackage{color}

\begin{document}
\title{Influence of scalar field in massive particle motion in JNW spacetime} 

\author{Bobur Turimov}\email{bturimov@astrin.uz}
\affiliation{Institute of Fundamental and Applied Research, National Research University TIIAME, Kori Niyoziy 39, Tashkent 100000, Uzbekistan}\affiliation{University of Tashkent for Applied Sciences, Str. Gavhar 1, Tashkent 100149, Uzbekistan}\affiliation{Shahrisabz State Pedagogical Institute, Shahrisabz Str. 10, Shahrisabz 181301, Uzbekistan}
\author{Akbar Davlataliev}\email{akbar@astrin.uz}
\affiliation{National University of Uzbekistan, Tashkent 100174, Uzbekistan} 
\author{Ahmadjon Abdujabbarov}
\email{ahmadjon@astrin.uz}\affiliation{Institute of Fundamental and Applied Research, National Research University TIIAME, Kori Niyoziy 39, Tashkent 100000, Uzbekistan}\affiliation{University of Tashkent for Applied Sciences, Str. Gavhar 1, Tashkent 100149, Uzbekistan}\affiliation{Shahrisabz State Pedagogical Institute, Shahrisabz Str. 10, Shahrisabz 181301, Uzbekistan}
\author{Bobomurat Ahmedov}\email{ahmedov@astrin.uz}
\affiliation{Institute of Fundamental and Applied Research, National Research University TIIAME, Kori Niyoziy 39, Tashkent 100000, Uzbekistan}\affiliation{National University of Uzbekistan, Tashkent 100174, Uzbekistan}\affiliation{Department of Physics and Mathematics, Uzbekistan Academy of Sciences, Y. Gulomov 70, Tashkent 100047, Uzbekistan} 

\date{\today}

\begin{abstract}

In this paper, we investigated the motion of massive particles in the presence of scalar and gravitational fields, particularly focusing on the Janis-Newman-Winicour (JNW) naked singularity solution. It is shown that the innermost stable circular orbit (ISCO) radius strongly depends on scalar coupling parameter. Additionally, we explored the radiation reaction effects on particle dynamics, incorporating a reaction term into the motion equations. Numerical simulations indicated minimal impact on particle trajectories from radiation reaction. We also examined the oscillatory motion of particles around compact objects in the JNW spacetime, focusing on radial and vertical oscillations. Our analysis indicated that the scalar field's coupling parameter and the spacetime deformation parameter $n$ significantly alter the fundamental frequencies of these oscillations. Furthermore, we studied quasi-periodic oscillations (QPOs) in X-ray binaries, using the relativistic precession (RP) model to analyze upper and lower frequency relationships. Our results indicated that increasing parameters ($n$ and $g_s$) shifts the frequency ratio of 3:2 QPOs closer to the naked singularity, with $n$ decreasing and $g_s$ increasing both frequencies. Finally, we analyzed QPO data from selected four X-ray binary systems using Markov Chain Monte Carlo (MCMC) analysis to constrain JNW parameters. Our findings provided insights into the mass, coupling and deformation parameter for each system, enhancing our understanding of compact object dynamics in strong gravitational fields. 
\end{abstract}
\maketitle

\section{Introduction}

There is enduring mystery within general relativity concerning the culmination of gravitational collapse of a massive astrophysical object. There is a belief that any complete gravitational collapse typically results in the formation of a Kerr black hole parameterized solely by its total mass and angular momentum. Other physical properties and 'hairs' at the initial stage of the massive body (such as  symmetries, the properties of the matter fields, etc.) are proposed to be emitted away as radiation~\cite{Penrose2002GReGr}. 
This conjecture has to be verified using the computational analysis taking into account all possible effects. Thus one must be careful with using this conjecture: it is not fully established yet that every gravitational collapse inevitably leads to black hole formation. It is worth to note that there are number of works showing that  under certain permissible initial conditions, gravitational collapses may result in the creation of naked singularities~\cite{
Joshi1993PhRvD,
Waugh1988PhRvD, 
Eardley1979PhRvD, 
Giambo2004GReGr, 
Joshi2002PhRvD.,
Harada1998PhRvD}.

In order to test the collapse mechanism and the product at final stage one needs to explore the the observational differences between black holes and naked singularities, assuming that they have been formed by some mechanism. At the present we have an access to big data of electromagnetic signals from astrophysical objects which can be used to  understand the nature of compact objects at the galactic centres or in the X-ray binaries. Particularly,  In Refs.~\cite{Kovacs2018PhRvD,
Blaschke2016PhRvD,
Stuchlik2010CQGra,
Bambi2013PhRvD,
Joshi2011CQGra,
Pugliese2011PhRvD,Pradhan2011PhLA.}. Authors have explored the accretion disks properties around naked singularity. The studies described in Refs.~\cite{
Hioki2009PhRvD,
Yang:2015hwf,
Takahashi2004IAUS,
Gyulchev2008PhRvD,
Virbhadra2002PhRvD, 
Virbhadra1998A&A,
Virbhadra2008PhRvD} 
have shown that black holes and naked singularities represent different properties in order to differentiate between them using gravitational lensing\cite{Deliyski2024arXiv,Virbhadra1998A&A,Chen2024EPJC,Chauvineau2022PhRvD,Gyulchev2020EPJC}. Energetic processes  and collision of particles may be used to distinguish the black holes from naked singularities~\cite{Patil:2011yb}. 

Particle motion in curved spacetime is a core concept in general relativity, explaining how matter and energy shape spacetime and how this curvature impacts the motion of objects. In such a curved spacetime, particles follow paths known as geodesics, governed by the geodesic equation. However, when an external electromagnetic field acts on a charged particle, it accelerates and emits electromagnetic radiation, which is described by a non-geodesic equation including interaction electromagnetic field and radiation bag reaction. Various types of electromagnetic radiation and their mathematical formulations can be found in \cite{Landau-Lifshitz2}. Additionally, considering particles in the presence of an external scalar field is both interesting and significant. The interaction between massive particles and scalar fields remains unclear, yet from a theoretical standpoint, it is crucial for understanding astrophysical processes near compact objects like black holes and neutron stars. This problem has been explored in Ref. \cite{Noble2021NJP} including self-force in the presence of the external scalar field. Certain aspects of this self-force can be interpreted through spacetime geometry, providing insight into the paradox of a particle radiating without experiencing a self-force. Acceleration of particle by black hole in the presence of the scalar field has investigated in \cite{Zaslavskii2017IJMPD}. The acceleration of particles by black holes in the presence of a scalar field has been examined in \cite{Zaslavskii2017IJMPD}. The interaction between scalar fields and massive particles was also introduced by Misner et al. \cite{Misner1972PRL} and by Breuer et al. \cite{Breuer1973PRD} to describe scalar perturbations or the so-called geodesic synchrotron radiation in Schwarzschild spacetime.

Here we plan to study the particle dynamics and quasiperiodic oscillations (QPO) around the Janis-Newman-Winicour (JNW) naked singularity. JNW naked singularity represents an exact solution of the Einstein’s equations with a massless scalar field~\cite{Janis:1968zz}. 
With the different parametrization the similar solution has been obtained by Fisher in  Ref.~\cite{Fisher1999gr.qc}
The stability of Fisher's solution has been studied in Ref.~\cite{Bronnikov:1979uz}

It is worth to note that dynamics of massive test particles around black holes represent a great interest in Astrophysics~\cite{5NA}. Particularly, one may use circular geodesics in order to describe the accretion processes around compact objects. A huge number of works are devoted to explore the particle dynamics and geodesics around compact object in different gravity models (see, e.g. Refs.~\cite{7,8,9,10,11,Hakimov17,12,13,14,Narzilloev21c,Narzilloev21d,Narzilloev22b,Narzilloev22a,Narzilloev23,Narzilloev22c,Davlataliev2023pdu,Narzilloev2023b,Narzilloev2023c,Narzilloev2023d,Narzilloev2023f,Narzilloev2023g,Narzilloev2023h,Rayimbaev:2023kxq,Abdulxamidov:2023jfq,Kurbonov:2023uyr,Abdulxamidov:2022ofi,Abdulkhamidov:2024lvp,Rayimbaev:2022pzr,Rayimbaev:2022hrn,Ladino:2023zdn,Rayimbaev:2023jmi,Davlataliev:2024ekv,16,17,18,19,2a,3a,Davlataliev:2024wdd, o1,o2,o3,o4, 6NA}). Furthermore, one may use the analysis of the geodesics to get constraints on different spacetime properties within different gravity theories~\cite{8NA,9NA,10NA}. 

One of the interesting features of the particle dynamics around compact objects is related to the observed QPOs from microquasars in  X-ray band \cite{4a}. The analysis QPOs may be considered as a useful tool to test the gravity theories in strong field regime. Particularly it allows to get the values of mass and spin of the black holes~\cite{N1,N2}. In order to describe QPOs various theoretical models have been proposed, namely  the disc-seismic models,  the warped disk models, the hot-spot models, the resonance models, etc. The different models and observations have been used to test and get constraints on different parameters of gravity theories~\cite{Rezzolla03,TKSS,SKT,SKT2,IH1,Re12,R13,Re14,Davlataliev:2024yle,Davlataliev:2024smq}.  %
In the present research paper, we aim to investigate particle dynamics in the vicinity of the JNW wormhole also known as JNW naked singularity which is the exact solution of Einstein-scalar field equation. We also plan to apply the results to study the QPOs around JNW compact objects.

The paper is organized as follows: In Sec.\ref{Sec:1} we briefly describe the Einstein-scalar field-massive particle system and derive the equation of motion for the whole system. In Sec.\ref{Sec:2} we consider the circular motion of massive particles in the presence of the scalar field in the JNW spacetime. In Sec.\ref{Sec:3} we study the circular motion of massive particles including radiation reaction term and present particle trajectories. In Sec.\ref{Sec:4}, we consider oscillatory motion of massive particle near the stable circular orbit in the JNW spacetime. As an astrophysical consequence we study application of the oscillatory motion of massive particle twin-peak QPO in Sec.\ref{Sec:5} and using these results obtain the constrains for two main parameters mass of the central object, scalar parameter and coupling constant in Sec.\ref{Sec:6}.  Finally, in Sec.\ref{Sec:7} we summarize the finding results. Throughout the paper, we use geometric unit $G=c=\hbar=1$.

\section{Einstein-scalar field-massive particle system\label{Sec:1}}

The action for the free relativistic particle of mass $m$ can be expressed as \cite{Landau1980}
\begin{align}
S=-\int mds\ ,    
\end{align}
while in the presence of the external scalar field $\varphi$, it reads \cite{Breuer1973PRD}
\begin{align}\label{action0}
S=-\int m_* ds\ , \qquad m_*=m(1+g_s\varphi)\ ,    
\end{align}
where $m_*$ is the effective mass of the test particle in the presence of a scalar field, and $g_s$ is a dimensionless coupling constant between scalar and massive particle. Notice that in Ref.~\cite{Misner1972PRL,Breuer1973PRD} that the action for massive particle and scalar field can be expressed as follows:
\begin{align}\nonumber
S[x^\mu, \varphi]=&-\frac{1}{8\pi}\int d^4x\, \partial_\mu\varphi\partial^\mu\varphi\\&-m\int ds(1+g_s\varphi)\sqrt{-u_\mu u^\mu}\ .  \label{action1}        
\end{align}
One has to emphasise that the action for massive particle in equation \eqref{action1} is also valid in curved spacetime. The generalized form of the action for the Einstein-scalar field system, including massive particles, is given as 
\begin{align}\label{action}\nonumber
S[g_{\mu\nu}, x^\mu, \varphi]&=\frac{1}{16\pi}\int d^4x\sqrt{-g}(R-2\nabla_\mu\varphi\nabla^\mu\varphi)\\&-m\int ds(1+g_s\varphi)\sqrt{-u_\mu u^\mu}\ ,   
\end{align}
where $R$ is the Ricci scalar, $\nabla_\mu$ stands for a covariant derivative from a scalar field and $u^\mu={\dot x}^\mu$ is the four-velocity of particle normalized as $u_\mu u^\mu=-1$. Hereafter minimizing the action \eqref{action}, equations of motion for whole system, namely Einstein field equations, Klein-Gordon equation and geodesic equation, can be derived as
\begin{align}\nonumber
R_{\mu\nu}&-\frac{1}{2}g_{\mu\nu}R=2\nabla_\mu\varphi\nabla_\nu\varphi-g_{\mu\nu}\nabla_\lambda\varphi\nabla^\lambda\varphi\\&
+\frac{8\pi m}{\sqrt{-g}}\int ds(1+g_s\varphi)\delta^{(4)}[x-z(s)]u_\mu u_\nu\ ,\\
&\nabla_\mu\nabla^\mu\varphi=\frac{4\pi mg_s}{\sqrt{-g}}\int ds\delta^{(4)}[x-z(s)]\ ,\\
&u^\mu\nabla_\mu u^\nu=\frac{g_s}{1+g_s\varphi}(g^{\mu\nu}+u^\mu u^\nu)\nabla_\mu\varphi\ .
\end{align}

For simplicity, we assume that the contribution of particle motion in the background spacetime and scalar field is negligibly small, and it does not alter the background geometry or the configuration of the scalar field. Consequently, equations of motion for the system reduces to
\begin{align}\label{eq1}
&R_{\mu\nu}=2\nabla_\mu\varphi\nabla_\nu\varphi\ ,\\\label{eq2}
&\nabla_\mu\nabla^\mu\varphi=0\ ,\\\label{eq3}
&u^\mu\nabla_\mu u^\nu=\frac{g_s}{1+g_s\varphi}(g^{\mu\nu}+u^\mu u^\nu)\nabla_\mu\varphi\ .
\end{align}
which are rather simply system of equations. In next section, we will discuss the explicit solution of above equations.

\section{Background spacetime and particle dynamics\label{Sec:2}}

In this section, we will explore the motion of a massive particle within scalar and gravitational fields. The system of equations \eqref{eq1} and \eqref{eq2} is straightforward, allowing for an exact analytical solution. One of the simplest solutions to these Einstein-scalar field equations is represented by the Janis-Newman-Winicour naked singularity. The corresponding line element is given as \cite{Janis1968PRL}
\begin{align}\nonumber\label{metric}
ds^2=&-f^ndt^2+f^{-n}dr^2+f^{1-n}r^2(d\theta^2+\sin^2\theta d\phi^2)\ , \\
\varphi(r)&=\frac{\sqrt{1-n^2}}{2}\ln f\ , \qquad f=1-\frac{2M}{nr}\ ,
\end{align}
where $M$ is the mass of the gravitational object, $n$ is parameter of the scalar field. The singularity of the spacetime is located at $r_*=2M/n$. The components Ricci tensor are expressed by
\begin{align}
R_{\mu\nu}=\frac{2(1-n^2)M^2}{f^2n^2r^4}\left(
\begin{array}{cccc}
 0 & 0 & 0 & 0 \\
 0 & 1 & 0 & 0 \\
 0 & 0 & 0 & 0 \\
 0 & 0 & 0 & 0 \\
\end{array}
\right)\ .    
\end{align}
When $n=1$, the solution \eqref{metric} reduces to the Schwarzschild spacetime, characterized by a horizon at $r=2M$ and satisfying $R_{\mu\nu}=0$.

The equation of motion for a massive particle in equation \eqref{eq3} in the background spacetime \eqref{metric}, can be explicitly written as:
\begin{align}
&{\ddot t}+\frac{2M{\dot t}{\dot r}}{fr^2}\left(1-B\right)=0\ ,\\\nonumber
&{\ddot r}+\frac{Mf^n}{fr^2}\left(f^n{\dot t}^2-2B\right) - \frac{M{\dot r}^2}{fr^2}\left(1+2B\right)\\&+\left(\frac{M(n+1)}{n}-r\right)\left({\dot\theta}^2+\sin^2\theta{\dot\phi}^2\right)=0\ ,\\
&\ddot\theta-\frac{1}{2}\sin 2\theta{\dot\phi}^2-\frac{2M{\dot\theta}{\dot r}}{fr^2}\left(1+\frac{1}{n}-\frac{r}{M}+B\right)=0\ , \\
&{\ddot\phi}+2\cot\theta{\dot\theta}{\dot\phi}-\frac{2M{\dot r}{\dot\phi}}{fr^2}\left(1+\frac{1}{n}-\frac{r}{M}+B\right)=0\ ,
\end{align}
where $B$ is defined as 
$$
B=\frac{2g_s\sqrt{1-n^2}}{n(g_s\sqrt{1-n^2}\ln f+2)}\ .
$$
 
On the other hand, using the equation of motion \eqref{eq3}, the Lagrangian for a massive particle in the presence of the scalar field can be expressed as:
\begin{align}
L=\frac{1}{2}(1+g_s\varphi)g_{\mu\nu}u^\mu u^\nu \ ,   
\end{align}
and the four-momentum of the massive particle is 
\begin{align}
P_\mu=\frac{\partial L}{\partial u^\mu}=(1+g_s\varphi)g_{\mu\nu}u^\nu\ .    
\end{align}

Since the spacetime is independent of the coordinates $t$ and $\phi$, the corresponding components of the four-momentum, $P_t$ and $P_\phi$, are conserved. These correspond to two conserved quantities: the energy and angular momentum of the massive particle. The explicit expressions for the specific energy ${\cal E}$ and specific angular momentum ${\cal L}$ are:
\begin{align}
&{\cal E}=-\frac{P_t}{m}=-(1+g_s\varphi)g_{tt}u^t\ ,\\ &{\cal L}=\frac{P_\phi}{m}=(1+g_s\varphi)g_{\phi\phi}u^\phi \ .   
\end{align}

Using the normalization of the four-velocity of particle, one can obtain
\begin{align}\label{eom}
g_{rr}{\dot r}^2+g_{\theta\theta}{\dot\theta}^2+1+\frac{1}{(1+g_s\varphi)^2}\left(\frac{{\cal E}^2}{g_{tt}}+\frac{{\cal L}^2}{g_{\phi\phi}}\right)=0\ .    
\end{align}

Here we focus on finding the ISCO position of massive particle in the JNW spacetime in the presence of the scalar field. For qualitative analyses, let us consider motion of massive particle in the equatorial plane, (i.e. $\theta=\pi/2$ and $\dot\theta=0$). Hereafter performing simple algebraic manipulations equation \eqref{eom} can be rewritten as
\begin{align}
{\cal E}^2=(1+g_s\varphi)^2{\dot r}^2+U(r)\ ,
\end{align}
where
\begin{align}
U(r)=f^{n}(1+g_s\varphi)^2+\frac{{\cal L}^2}{r^2}f^{2n-1}\ ,    
\end{align}
is the effective potential for massive particle. 

The ISCO radius of massive test particle is a crucial concept in the study of the dynamics around compact astrophysical objects such as black holes and neutron stars. It represents the smallest radius at which a test particle can stably orbit the central object. Inside this radius, any perturbation will cause the particle to spiral inward, leading to eventual accretion onto the central body. For the Schwarzschild black hole, the ISCO radius is located at $r_{\rm ISCO}=6M$. In the case of a rotating Kerr black hole, the ISCO radius depends on the black hole's spin parameter and the direction of the particle's orbit. For prograde orbits (orbits in the direction of the black hole's spin), the ISCO radius decreases with increasing spin, potentially reaching as close as the event horizon for an extremely spinning black hole. Conversely, for retrograde orbits (orbits opposite to the direction of the black hole's spin), the ISCO radius increases with the black hole's spin. The ISCO radius is significant in astrophysics for several reasons: (i) Accretion Disk Dynamics: It defines the inner edge of the accretion disk around compact objects. The region inside the ISCO is dynamically unstable, causing material to rapidly spiral inwards and potentially emit significant amounts of radiation, (ii) Gravitational Wave Emission: For compact binary systems, the ISCO sets a natural frequency cutoff for gravitational wave emission, influencing the waveforms detected by observatories like LIGO and Virgo.  (iii) Astrophysical Observations: Observations of the X-ray spectra from accreting black holes can provide estimates of the ISCO radius, thereby offering insights into the spin and other properties of the black hole.

\begin{figure}
\centering\includegraphics[width=0.92\linewidth]{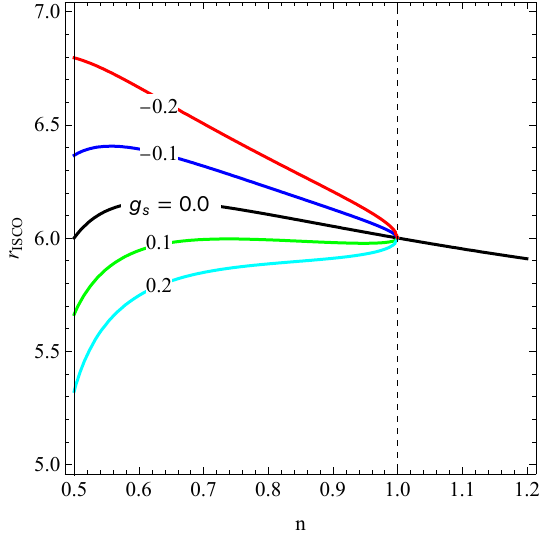}\caption{Dependence of the ISCO position of massive particle from the parameter $n$ for the different values of the coupling parameter.}\label{fig_ISCO}
\end{figure}

It is also interesting to probe how the scalar field influences to the ISCO position for massive particle in the JNW spacetime. Using the conditions ${\cal E}^2 = U(r)$ and $U'(r) = 0$, one obtains the critical specific energy and critical specific angular momentum as follows:
\begin{align}\label{E2}
&{\cal E}^2=\frac{f^n (1+g_s\varphi) \left((1+g_s\varphi) \left(2 f-(n-1) r f'\right)+2 g_s r f \varphi'\right)}{(1-2n)rf'+2f} \ ,
\\\label{L2}
&{\cal L}^2=\frac{r^3 f^{1-n} (1+g_s\varphi) \left(f'n(1+g_s\varphi )+2 g_s f \varphi'\right)}{(1-2n)rf'+2f} \ .
\end{align}
The stationary points of these expressions \eqref{E2} and \eqref{L2} represent the innermost stable circular orbit (ISCO). If one neglects the interaction term ($g_s = 0$), the ISCO radius is determined as:
\begin{align}
r_{\rm ISCO}=\left(3+\frac{1}{n}+\sqrt{5-\frac{1}{n^2}}\right)M\ .    
\end{align}
The ISCO radius is also influenced by various factors including the presence of additional forces or fields. For example, in the presence of an interaction parameter $ g_s $, the ISCO equation can become quite complex. Numerical calculations have shown that the ISCO radius for a massive particle decreases when $ g_s > 0 $ and increases when $ g_s < 0 $. This indicates that the nature of the interactions around the central object can significantly alter the orbital dynamics, providing a rich field of study in both theoretical and observational astrophysics. In Fig.\ref{fig_ISCO}, we show dependence of the ISCO position for massive particle from scalar parameter $n$ for different value of the interaction parameter $g_s$.

\section{Radiation Reaction\label{Sec:3}}

Since the four-acceleration of massive particles in the presence of a scalar field is non-zero, i.e., $w^\nu=u^\mu \nabla_\mu u^\nu \neq 0$, these particles can emit either electromagnetic or gravitational radiation. However, we currently lack understanding of the specific type of radiation emitted by test particles in this scenario. In the classical limit of this radiation, the coupling parameter $g_s$ is related to the Newtonian gravitational constant. Thus, this radiation can be considered analogous to gravitational radiation. The intensity of the radiation from particle in the presence of the scalar field is given by  
\begin{align}
I \sim w_\mu w^\mu=\frac{g_s^2}{(1+g_s\varphi)^2}(g^{\mu\nu}+u^\mu u^\nu)\nabla_\mu\varphi\nabla_\nu\varphi\ .    
\end{align}

Now we explore the dynamics of massive particles in the presence of a scalar field, incorporating a radiation reaction term. While the Lorentz-Abraham-Dirac (LAD) equation is widely recognized as the governing equation for radiation reaction in charged particles (see, for example, \cite{Poisson2004LRR, Tursunov2018ApJ}), our study focuses on its gravitational counterpart. In this context, the equation of motion \eqref{eom} can be modified as follows \cite{Noble2021NJP}:
\begin{align}\label{LAD}\nonumber
\frac{Du^\mu}{ds}&=\frac{g_s}{1+g_s\varphi}(g^{\mu\nu}+u^\mu u^\nu)\partial_\nu\varphi\\\nonumber&+\frac{1}{2}\tau_0\left(R^\mu_{~\nu}+u^\mu u_\lambda R^\lambda_{~\nu}\right)u^\nu\\&+\tau_0\left(\frac{D^2u^\mu}{ds^2}+u^\mu u_\lambda\frac{D^2u^\lambda}{ds^2}\right)\ ,
\end{align}
where $\tau_0$ stands for the damping time of gravitational radiation defined as $\tau_0<<\tau$, and this parameter serves as an expansion factor. 

Given that the final term of equation \eqref{LAD} is significantly smaller compared to the other terms, the Landau trick can be employed to expand the equation in terms of the damping time $\tau_0$. By differentiating equation \eqref{LAD} and neglecting the highest-order terms related to the damping time $\tau_0$, we obtain:
\begin{align}\nonumber\label{Expand}
&\frac{D^2u^\mu}{ds^2}=\frac{g_s}{(1+g_s\varphi)}(g^{\mu\nu}+u^\mu u^\nu)u^\alpha\partial_\alpha\partial_\nu\varphi\\&+\frac{g_s^2}{(1+g_s\varphi)^2}(g^{\nu\alpha}+u^\nu u^\alpha)u^\mu\partial_\alpha\varphi\partial_\nu\varphi+{\cal O}(\tau_0)\ ,
\end{align}

Substituting expression \eqref{Expand} into \eqref{LAD} and after simple algebraic manipulations, the equation of motion can be reformulated as:
\begin{align}\label{Rad}\nonumber
\frac{Du^\mu}{ds}&=\frac{g_s}{1+g_s\varphi}(g^{\mu\nu}+u^\mu u^\nu)(\partial_\nu\varphi+\tau_0u^\alpha\partial_\alpha\partial_\nu\varphi)\\&+\frac{1}{2}\tau_0\left(R^\mu_{~\nu}+u^\mu u_\lambda R^\lambda_{~\nu}\right)u^\nu\ .    
\end{align}

The main idea behind reducing equation \eqref{LAD} to the form of \eqref{Rad} is to simplify the system into a second-order differential equation for four coordinates $x^\mu$ instead of a third-order system. This reduces the necessity of finding twelve constants of motion, a primary challenge in particle motion in curved spacetime. By employing the Landau trick, the equation of motion can be expressed as a second-order system for the four coordinates $x^\mu$. While the analytical form of the equations for each coordinate in \eqref{Rad} is extensive, we refrain from reporting them in this letter. Nevertheless, careful numerical analyses for given initial conditions $x_0^\mu=(0,r_0,\theta_0,0)$ and $u_0^\mu=\left(\frac{\cal E}{f(r_0)^n}, 0, 0, \frac{{\cal L}f(r_0)^{n-1}}{(r_0 \sin \theta_0)^2}\right)$ enable the determination of each coordinate as a function of the affine parameter, i.e., $x^\mu = x^\mu(s)$. 

\begin{figure}
\centering\includegraphics[width=0.92\linewidth]{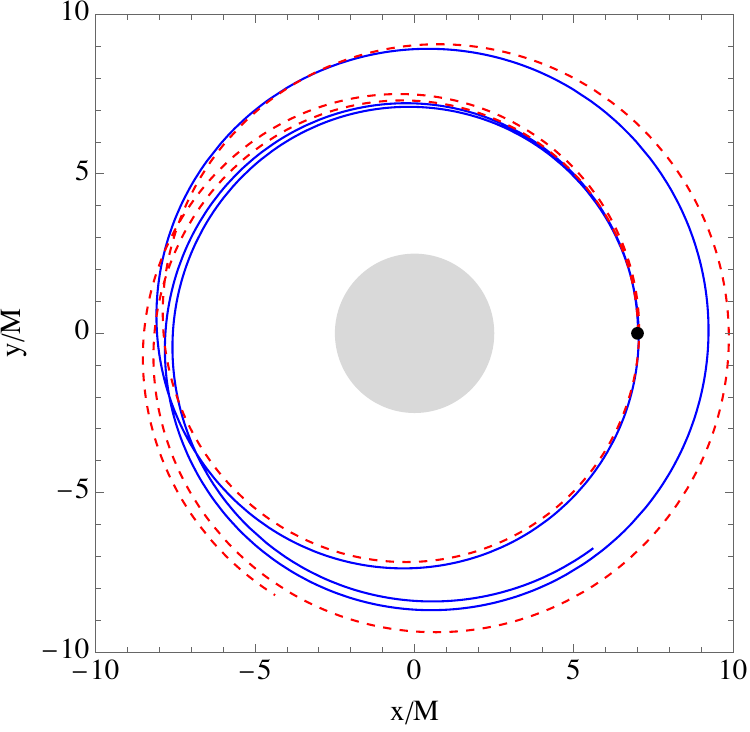}
\caption{Particle trajectories near the JNW naked singularity without ($\tau_0=0$, solid blue line) and with ($\tau_0=1.0$, dashed red line) radiation reaction term for the particular choice of parameters $\text{g}_s=0.03$ and $n=0.8$}\label{trajectories}
\end{figure}

\begin{figure}   
\centering\includegraphics[width=0.92\linewidth]{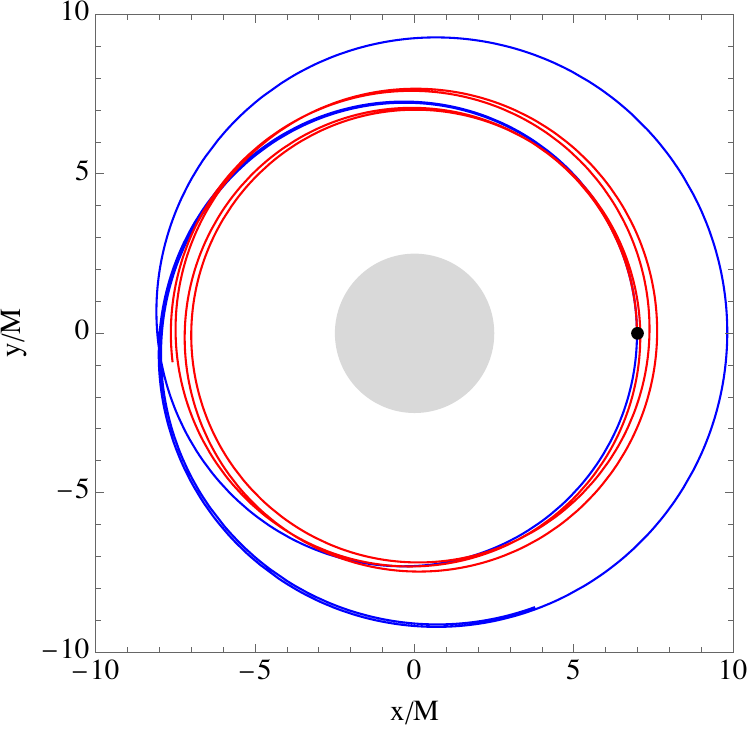}
\caption{Particle trajectories near the JNW naked singularity the particular choice of parameters. Blue solid line represents for $g_s=0.03$ and $\tau_0=0$ s and red solid line represents for $g_s=0$ and $\tau_0=0.5$ s.}\label{mimic}
\end{figure}

In Fig. \ref{trajectories}, we show the trajectory of a massive particle in the presence of a scalar field, including the radiation reaction term. Using numerical calculations, we have demonstrated that the contribution due to the radiation reaction is not significant. Therefore, to observe the effect of the radiation reaction term in circular motion, we set the damping time to $\tau_0 = 0$ s and $\tau_0 = 0.5$ s. This implies that the damping time is much shorter than 1 s, and the particle orbits many more times than shown in Fig. \ref{trajectories} around the compact object described by JNW spacetime before eventually escaping due to the radiation reaction force. It is shown that unlike a black hole \cite{Turimov2023PDU}, a naked singularity generates a repulsive effect in the radiation reaction due to the scalar field. In Fig. \ref{mimic}, we demonstrate the trajectory of a massive particle in two scenarios: (i) considering effect from the scalar field but without the radiation reaction term, and (ii) neglecting effect from the scalar field but including the radiation reaction term. This result show that indeed the scalar field generates attractive effect however radiation reaction repulsive effect in the motion of massive particle orbiting around the naked singularity.

\section{Oscillatory motion \label{Sec:4}}

Oscillatory motion of test particles around gravitational compact objects is a key aspect of relativistic astrophysics. This motion can be categorized into radial and vertical oscillations, and it provides crucial insights into the dynamics and structure of the spacetime around these objects. (i) Radial oscillations refer to the back-and-forth motion of a particle in the radial direction, i.e., towards and away from the central object. These oscillations occur around stable circular orbits. The stability and frequency of these oscillations are influenced by the spacetime geometry and the properties of the massive object. For the Schwarzschild black hole, the radial oscillation frequency depends on the mass of the black hole and the radial distance from it. Closer to the black hole, the stronger gravitational pull leads to higher oscillation frequencies. (ii) Vertical oscillations, also known as latitudinal or epicyclic oscillations, refer to the motion of a particle perpendicular to the plane of the orbit. These oscillations provide information about the vertical stability of the orbit and are influenced by the angular momentum and spin of the central object. The vertical epicyclic frequency is the frequency at which a particle oscillates vertically around the equatorial plane of the central object. These two kind frequencies are related to the orbital frequency also known as Keplerian frequency. All together they are called fundamental frequencies and in the Schwarzschild black hole, these fundamental frequencies are expressed as $\Omega_r=\Omega_K\sqrt{1-6M/r}$, and $\Omega_\theta=\Omega_K=\sqrt{M/r^3}$, where $\Omega_K$ is Keplerian frequency in the Schwarzschild spacetime.

Here we will focus on oscillatory motion of massive particle near the stable circular orbit in the JNW spacetime in the presence of the scalar field. In this case four-velocity of particle takes a form: $u^\mu={\dot t}(1,0,0,\Omega)$, where $\Omega=d\phi/dt$ is the angular velocity of the particle. The radial equation reduces to
\begin{align}\label{W}
\Omega^2=-\frac{(1+g_s\varphi)\partial_r g_{tt}-2g_s g_{tt}\varphi'(r)}{(1+g_s\varphi)\partial_r g_{\phi\phi}-2g_s g_{\phi\phi}\varphi'(r)}\ .
\end{align}
Notice that in absence of the scalar field the explicit expression for the angular velocity of massive particle in JNW spacetime reduces to $\Omega=\Omega_Kf^{n-1/2}\left[1-M(1+n)/nr\right]^{-1}$. Consequently, expressions for the specific energy and angular momentum of massive particle yield  
\begin{align}
&{\cal E}=-(1+g_s\varphi)g_{tt}{\dot t}=-\frac{(1+g_s\varphi)g_{tt}}{\sqrt{-g_{tt}-\Omega^2g_{\phi\phi}}}\ ,
\\ 
&{\cal L}=(1+g_s\varphi)g_{\phi\phi}\Omega{\dot t}=\frac{(1+g_s\varphi)g_{\phi\phi}\Omega}{\sqrt{-g_{tt}-\Omega^2g_{\phi\phi}}}\ .   
\end{align}

By using the normalization of the four-velocity of massive particle, one can obtain
\begin{align}\label{2D}
g_{rr}{\dot r}^2+g_{\theta\theta}{\dot\theta}^2+V(r,\theta)=0\ ,    
\end{align}
where $V(r,\theta)$ is the effective potential explicitly given in equation \eqref{eom}. From the expression \eqref{2D}  equation for $2D$-harmonic oscillator can be derived after expanding it up to leading order as follows
\begin{align}\nonumber
&g_{rr}({\dot r}_0+\dot {\delta r})^2+g_{\theta\theta}({\dot\theta}_0+{\dot\delta\theta})^2+V(r_0,\theta_0)\\\nonumber&+\partial_rV(r_0,\theta_0)\delta r+\partial_\theta V(r_0,\theta_0)\delta\theta+\partial_r\partial_\theta V(r_0,\theta_0)\delta r\delta\theta\\&+\frac{1}{2}\partial_r^2V(r_0,\theta_0)\delta r^2+\frac{1}{2}\partial_\theta^2V(r_0,\theta_0)\delta\theta^2+...=0\ .    
\end{align}
Consequently, equations of oscillatory motion for radial and vertical displacements $\delta r$ and $\delta\theta$ from the stationary point read as follows: 
\begin{align}\nonumber
&g_{rr}\left(\frac{d^2}{dt^2}+\Omega_r^2\right){\delta r}+g_{\theta\theta}\left(\frac{d^2}{dt^2}+\Omega_\theta^2\right)\delta\theta=0\ , 
\end{align}
where
\begin{align}
&\Omega_r^2=\frac{1}{2g_{rr}{\dot t}^2}\partial_r^2V=\frac{-g_{tt}-\Omega_{\phi}^2 g_{\phi\phi}}{2g_{rr}}\partial_r^2V
\\\label{Wq}
&\Omega_\theta^2=\frac{1}{2g_{rr}{\dot t}^2}\partial_\theta^2V=\frac{-g_{tt}-\Omega_{\phi}^2 g_{\phi\phi}}{2g_{\theta\theta}}\partial_\theta^2V\ .
\end{align}
Our analysis showed that vertical frequency in \eqref{Wq} is the same as orbital angular velocity in \eqref{W} (i.e. $\Omega_\theta=\Omega$). We express all fundamental frequencies in Hz throughout our analysis. Therefore we restore fundamental constants in all frequencies as follows 
$$\nu_i = \frac{1}{2\pi}\frac{c^3}{G M} \Omega_i$$
where $c= 3\times 10^{10}{\rm cm s^{-1}}$  is the speed of light in a vacuum and $G=6.67\times 10^{-8}{\rm cm^3 g^{-1} s^{-2}}$  is the Newtonian gravitational constant.

\begin{figure*}\centering
\includegraphics[width=0.45\linewidth]{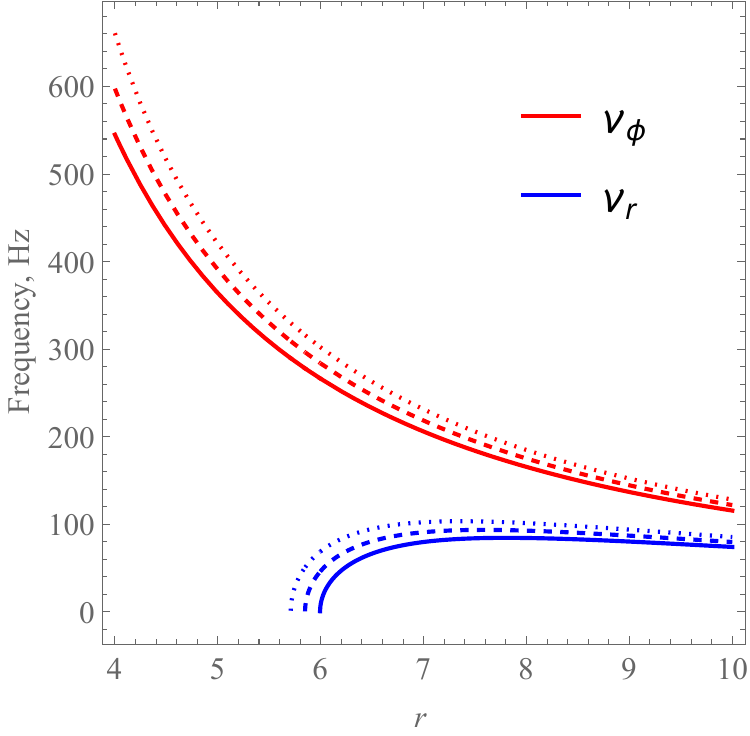}
\includegraphics[width=0.45\linewidth]{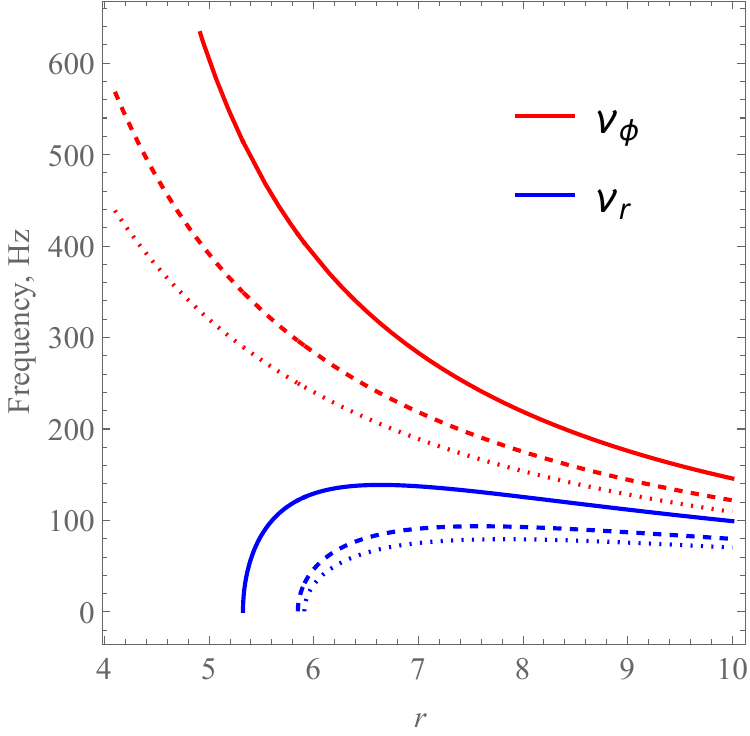}\caption{(Left panel) Radial dependence of fundamental frequencies of massive particle for different values of the coupling parameter ($g_s=0.1$ - solid line, $g_s = 0.2 $ - dashed line, and $ g_s = 0.3$ -dotted line) for fixed $n=0.7$. (Right panel) Radial dependence of fundamental frequencies of massive particle for different values of the $n$ parameter ($n=0.5$ - solid line, $n= 0.7$ - dashed line, and $n=0.9$ -dotted line) for fixed $g_s=0.2$. In both cases, the mass of the central object is taken as $M = 10M_\odot$.}\label{freq}
\end{figure*}

The radial dependence of fundamental frequencies is illustrated in Fig. \ref{freq} with specific parameter settings. The left panel displays plots for different values of the coupling parameters with a fixed value of the $ n $ parameter, while the right panel shows similar plots for varying values of the $ n $ parameter with a fixed coupling parameter $ g_s $. It can be observed that the fundamental frequencies increase due to the interaction of the massive particle with the scalar field, but they decrease with an increase in the $ n $ parameter.

\section{QUASI-PERIODIC OSCILLATION\label{Sec:5}}

The combined effect of orbital, radial and vertical oscillations can lead to complex trajectories for test particles. These oscillations are crucial for understanding phenomena such as quasi-periodic oscillations (QPOs) observed in X-ray binaries. QPOs are believed to be related to the oscillatory motion of matter in the accretion disk around a black hole or neutron star. The oscillatory motion affects the stability and structure of accretion disks. Understanding these oscillations helps in modeling the emission spectra and variability of the disks. Observing the frequencies and modes of oscillations can provide insights into the properties of black holes and neutron stars, such as their mass, spin, and the geometry of the surrounding spacetime. 

Astrophysical compact objects like black holes do not directly emit electromagnetic radiation. However, their strong gravitational pull distorts spacetime, significantly affecting the motion of nearby matter. This gravitational influence can lead to the formation of an accretion disk a rotating mass of gas, dust, and other debris caught in the black hole's gravitational field. As the material within the accretion disk orbits the black hole, friction and other forces generate extreme heat and pressure, causing the disk to emit large amounts of electromagnetic radiation across various wavelengths. Thus, while black holes themselves cannot be observed directly, the luminous emissions from their accretion disks offer valuable insights into their presence and properties in the universe.

The Relativistic Precession (RP) model offers a theoretical framework to explain the occurrence of quasi-periodic oscillations (QPOs), attributing them to the quasi-harmonic oscillations of charged particles as they move radially and angularly around black holes and wormholes. This model posits a direct connection between the dynamics of these charged particles and the manifestation of QPO phenomena, providing valuable insights into the mechanisms driving oscillatory behavior in astrophysical systems with strong gravitational fields.

Within the RP model, the interpretation of twin-peaked QPOs is explained as follows: the higher frequency corresponds to the orbital frequency of the particle, denoted as $ \nu_U = \nu_\phi $, while the lower frequency corresponds to the difference between the orbital frequency and the radial oscillation frequency, expressed as $ \nu_L = \nu_\phi - \nu_r $. In other words, for twin-peaked QPOs, the upper frequency reflects the particle's orbital frequency, and the lower frequency is obtained by subtracting the radial oscillation frequency from the orbital frequency. This provides a clear understanding of the frequency components observed in twin-peaked QPO phenomena within the RP model, enabling insightful analysis of astrophysical systems.

\begin{figure*}
    \centering
    \includegraphics[width=0.45\linewidth]{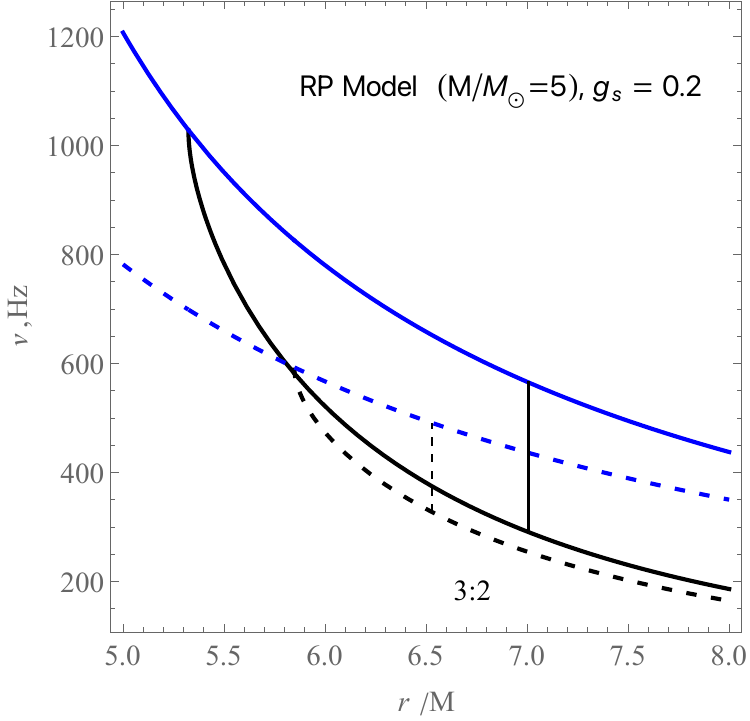}
    \includegraphics[width=0.45\linewidth]{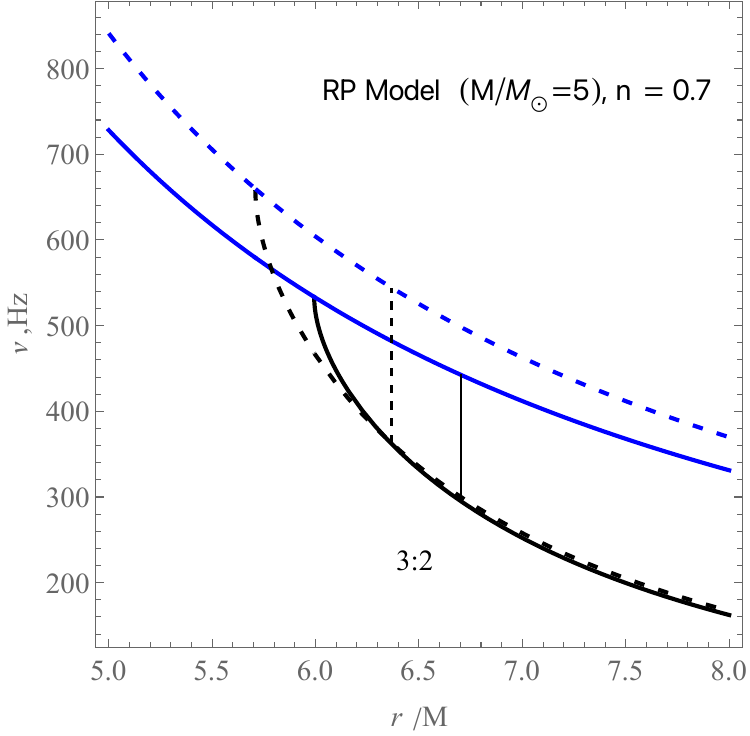}
    \caption{The radial dependence of upper and lower frequencies in RP model that are observed at ratio of 3:2. }
    \label{n32}
\end{figure*}

Figure \ref{n32} illustrates the radial dependence of the upper and lower frequencies of twin QPOs within the RP model, shown by blue and black lines, respectively. In the left panel, the solid line represents $ n = 0.5 $, while the dashed line represents $ n = 0.7 $ with a fixed coupling parameter $ g_s = 0.2 $. In the right panel, the solid line represents $ g_s = 0.1 $, while the dashed line represents $ g_s = 0.3 $ with a fixed $ n = 0.7 $. It is evident from the figures that the radial position at which the frequency ratio of 3:2 QPOs is observed shifts slightly toward the naked singularity as the two main parameters, $ n $ and $ g_s $, are increased.

\begin{figure*}
    \centering
    \includegraphics[width=0.45\linewidth]{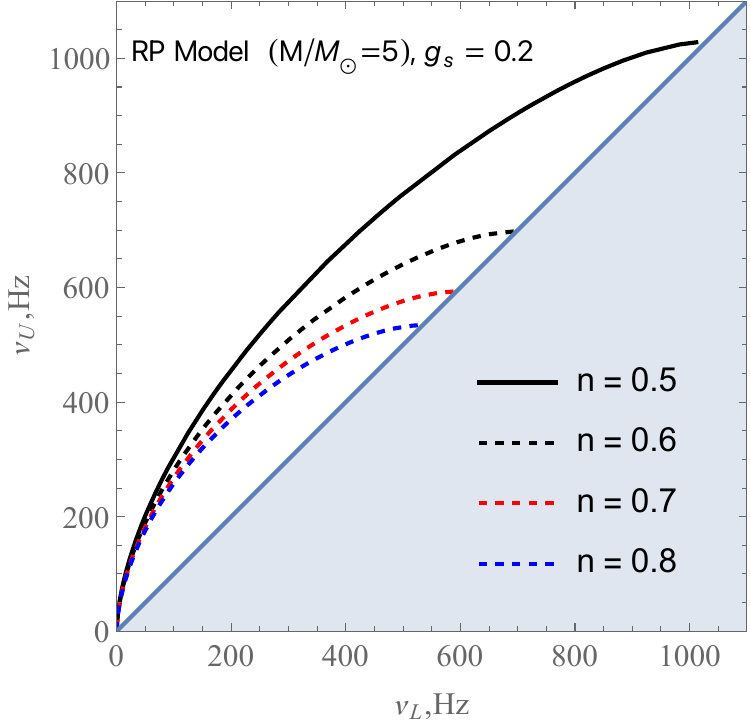}
    \includegraphics[width=0.45\linewidth]{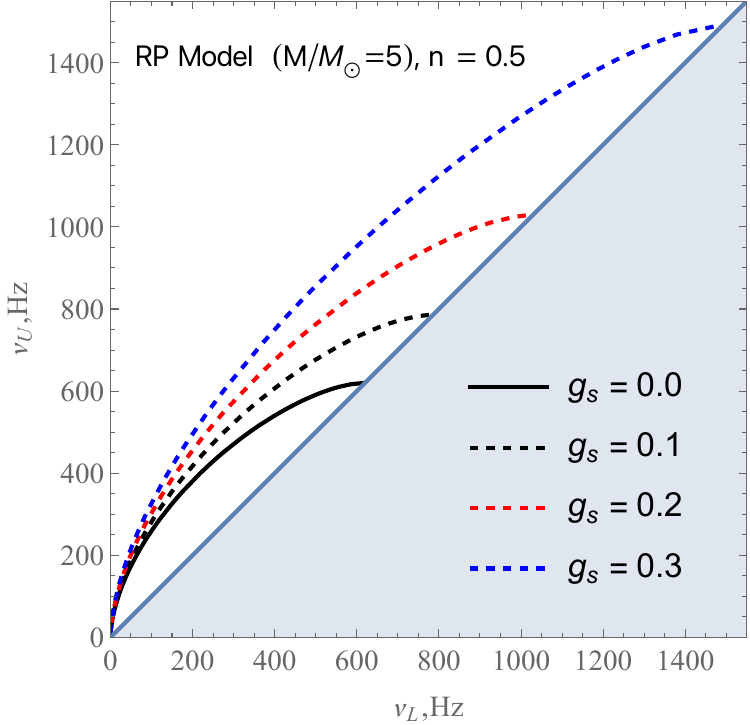}
    \caption{Relationships between the upper and lower peak frequencies of twin-peak QPOs in the RP model with mass $ M = 5 M_\odot $}
    \label{RPmodel}
\end{figure*}

In Fig. \ref{RPmodel}, we examine the relationships between the upper and lower frequencies of twin peak QPOs for various $ n $ and $ g_s $ values. The analysis shows that as $ n $ increases, both the upper and lower frequencies decrease. Conversely, the effect of $ g_s $ is opposite, leading to an increase in these frequencies. As can be seen from Fig. \ref{RPmodel}, the frequencies of the twin peak QPOs are not observable in the shaded area in both plots because these correspond to distances inside the ISCO position.

\section{Constraints\label{Sec:6}}

In this section, we analyze four X-ray binary systems to constrain the parameters associated with our JNW by examining Quasi-Periodic Oscillations (QPOs) data from these systems. The celestial objects under study include XTE J1550-564, GRO J1655-40 and GRS J1915+105. We will ultimately present the best-fit values within the parameter space, obtained through Markov Chain Monte Carlo (MCMC) code analysis.

We used the Python library emcee \cite{2013PASP125306F} for MCMC analysis to derive constraints on the JNW parameters in our study, which is based on the relativistic precession (RP) model.

The posterior is defined as described in reference \cite{Liu:2023vfh}:

\begin{align}\label{Dist}
\mathcal{P}(\Theta|D,M) = \frac{P(D|\Theta,M)\pi(\Theta|M)}{P(D|M)} \quad  \ , 
\end{align}

Where:
- $ \pi(\Theta) $ is the prior.
- $ P(D|\Theta, M) $ is the likelihood.

The priors are Gaussian within specified boundaries, i.e.,

\begin{align}\label{pitheta}
\pi(\Theta_i) \sim \text{exp}\left( \frac{{1}}{{2\sigma_i^2}} (\Theta_i - \Theta_0)^2 \right) \ , 
\end{align}
Where $ \Theta_{\text{low},i} < \Theta_i < \Theta_{\text{high},i} $ for the parameters $ \Theta_i = [M,g_s,n,r/M] $, and $ \sigma_i $ are their corresponding standard deviations. We adopt the prior values of the parameters for the JNW as shown in Table \ref{tablemean}.

Considering the upper and lower frequency data obtained in Section \ref{Sec:5}, our MCMC analysis incorporates two separate datasets. The core of this analysis is the likelihood function, denoted as $ \mathcal{L} $, which is expressed as follows:
\begin{align}\label{LogL}
\log \mathcal{L} = \log \mathcal{L}_{\text{up}} + \log \mathcal{L}_{\text{low}} \ , 
\end{align}

The logarithm of the likelihood function is the sum of the logarithms of the upper and lower frequency likelihoods. The logarithm of $\mathcal{L}_{\text{up}} $ represents the likelihood associated with the upper-frequency data, given by:

\begin{align}\label{LogL_up}
\log \mathcal{L}_{\text{up}} = -\frac{1}{{2}} \sum_{i} \left( \frac{{(\nu^{i}_{\text{up, obs}} - \nu^{i}_{\text{up, th}})^2}}{{(\sigma^{i}_{\text{up, obs}})^2}} \right)  \ , 
\end{align}
Meanwhile, $ \log \mathcal{L}_{\text{low}} $ represents the likelihood associated with the lower frequency data, given by:

\begin{align}\label{LogL_low}
\log \mathcal{L}_{\text{low}} = -\frac{1}{2} \sum_{i} \left( \frac{(\nu^{i}_{\text{low, obs}} - \nu^{i}_{\text{low, th}})^2}{(\sigma^{i}_{\text{low, obs}})^2} \right) \ , 
\end{align}

In these expressions, $ i $ ranges from 1 to an arbitrary integer, representing the number of measured upper and/or lower frequencies. $ \nu^i_{\text{up, obs}} $ and $ \nu^i_{\text{low, obs}} $ denote the observed results for the upper and lower frequencies, respectively, denoted as $ \nu_{\text{up}} $ and $ \nu_{\text{low}} $. Additionally, $ \nu^i_{\text{up, th}} $ and $ \nu^i_{\text{low, th}} $ correspond to their respective theoretical predictions. Furthermore, $ \sigma^i_{\text{up}} $ and $ \sigma^i_{\text{low}} $ represent the statistical uncertainties associated with these quantities.

\begin{figure*}
    \centering
    \includegraphics[width=0.45\linewidth]{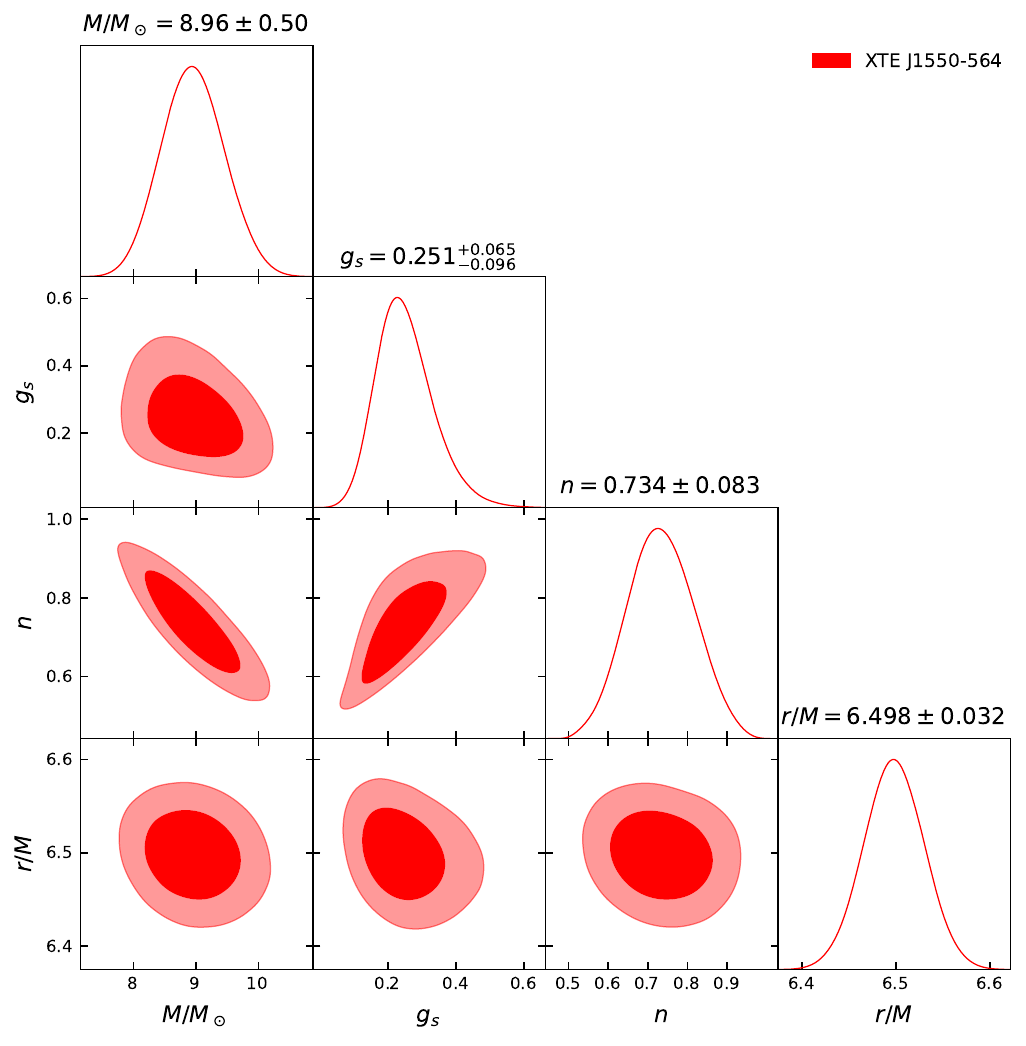}
    \includegraphics[width=0.45\linewidth]{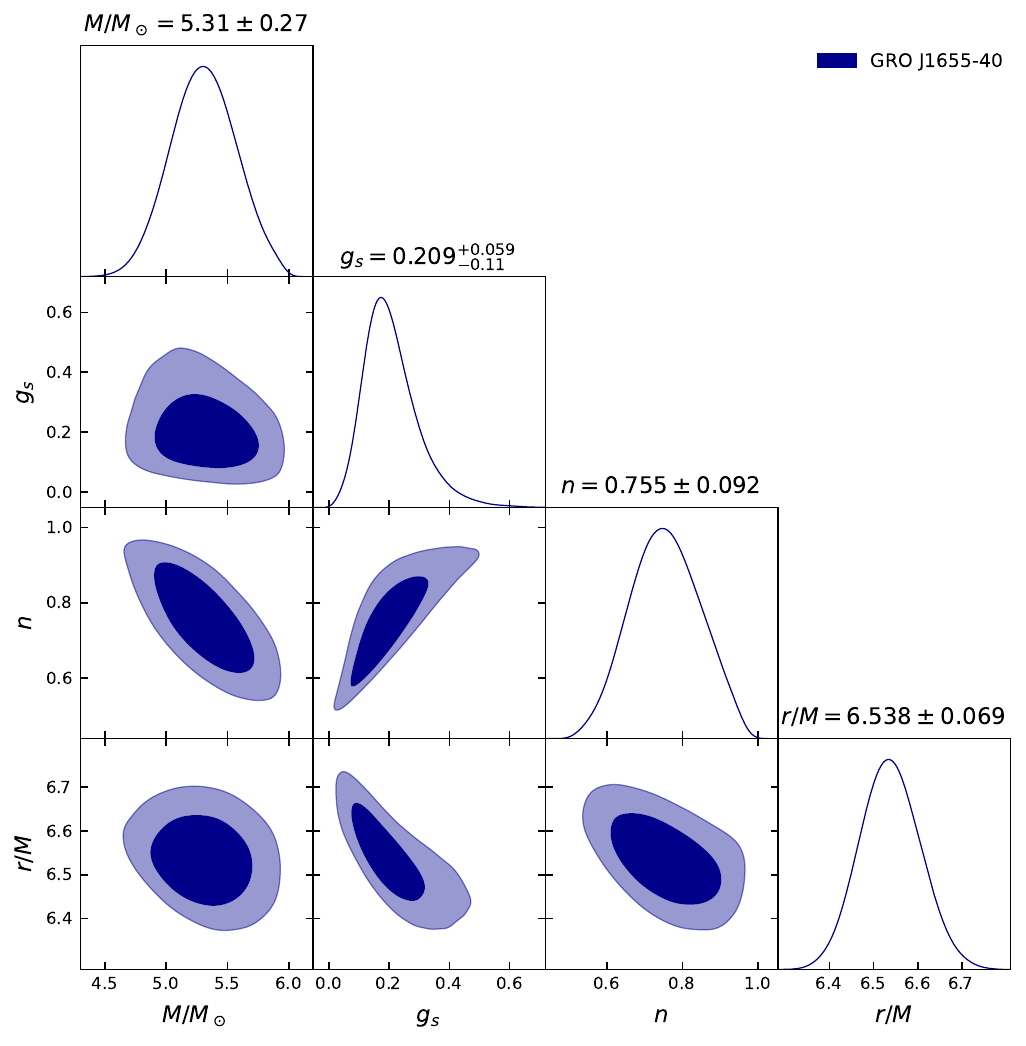}
     \includegraphics[width=0.45\linewidth]{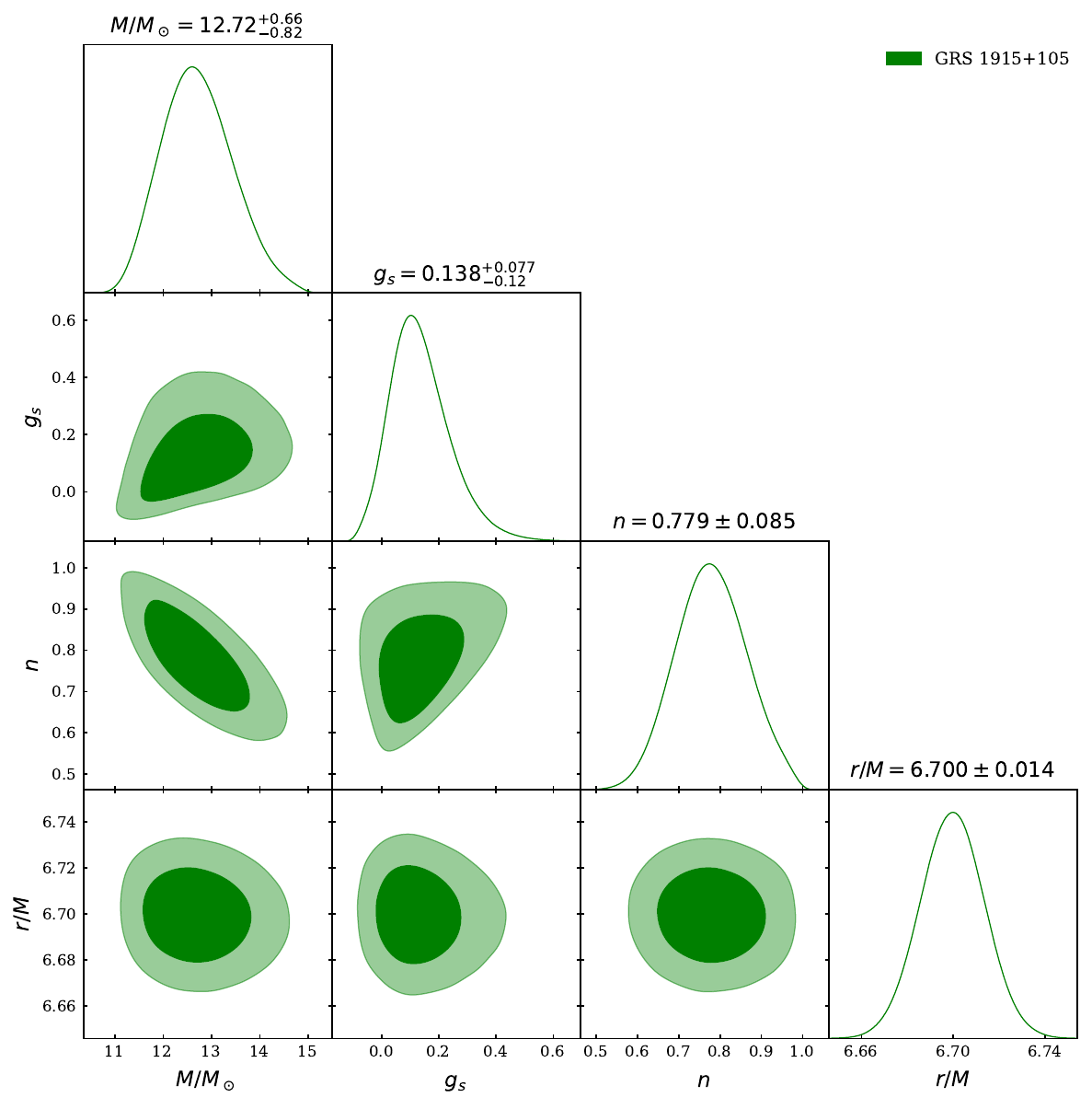}
    \caption{ Constraints on JNW mass, the $g_s$ and $n$ parameters in the microquasars XTE J1550-564(upper left),GRO J1655-40 (upper right) and GRS 1915+105 (down) using MCMC analysis.}
    \label{figmcmc}
\end{figure*}


\begin{center}
\begin{table}[h!]
\centering
\caption{The mass, orbital frequencies, periastron precision frequencies and nodal precision frequencies of QPOs from the X-ray binaries selected for analysis.} 
\begin{tabular}{||c c c c c||} 
 \hline
 & XTE J1550-564 & GRO J1655-40 &  GRS 1915+105 &\\ [0.6ex] 
 \hline\hline
 $M(M_{\odot})$ & 9.1 $\pm$ 0.61\cite{Remillard:2002cy,Orosz:2011ki}& 5.4$\pm$0.3\cite{Motta:2013wga} & $12.4^{+2.0}_{-1.8}$ \cite{Remillard:2006fc}&\\ 
 $\nu_{up}$(Hz) & 276 $\pm$ 3\cite{Remillard:2002cy} & 441$\pm$ 2\cite{Motta:2013wga} &  168 $\pm$ 3 \cite{Remillard:2006fc}&  \\
 $\nu_{low}$(Hz)& 184 $\pm$ 5 \cite{Remillard:2002cy} & 298 $\pm$ 4\cite{Motta:2013wga} & 113 $\pm$ 5 \cite{Remillard:2006fc}& \\[1ex] 
 \hline
\end{tabular}
\label{table:1}
\end{table}
\begin{table}[h!]
\centering
\caption{The Gaussian prior of the JNW derived from QPOs for the selected X-ray sources.
} 
\begin{tabular}{||c c c c c||} 
 \hline
 & XTE J1550-564 & GRO J1655-40 &GRS 1915+105 &\\ [0.6ex] 
 \hline\hline
 $M(M_{\odot})$ & $9.1\pm0.6$ &$5.4\pm0.3$ &$12.75\pm1.25$&\\ 
 $g_s$ & $0.25\pm0.15$& $0.25\pm0.2$&$0.225\pm0.225$&\\
 $n$ & $0.715\pm0.175$& $0.7225\pm0.1975$&$0.75\pm0.1$& \\$r/M$&$6.5\pm 0.033$&$6.56\pm0.08$&$6.7\pm0.014$&\\[1ex] 
 \hline
\end{tabular}
\label{tablemean}
\end{table}
\begin{table}[h!]
\centering
\caption{The best-fit parameter values resembling those of JNW, deduced from the Quasi-Periodic Oscillations (QPOs) for the chosen X-ray sources.} 
\begin{tabular}{||c c c c c||} 
 \hline
 & XTE J1550-564 & GRO J1655-40 & GRS 1915+105 &\\ [0.6ex] 
 \hline\hline
 $M(M_{\odot})$ & $8.96\pm0.50$ &$5.31\pm0.27$ &$12.72^{+0.66}_{-0.82}$&\\ $g_s$&$0.251^{+0.065}_{-0.096}$&$0.209^{+0.059}_{-0.11}$&$0.138^{+0.077}_{-0.12}$&\\
 $n$ & $0.734\pm0.083$& $0.755\pm0.092$&$0.779^\pm0.085$& \\$r/M$&$6.498\pm0.032$&$6.538\pm0.069$&$6.700\pm0.014$& \\[1ex] 
 \hline
\end{tabular}
\label{table2}
\end{table}

\end{center}

Using MCMC analyses at $1\sigma$ confidence level in Fig.\ref{figmcmc}, we show the corner plots three different selected sources as mentioned before for the best-fit parameter values ($M, g_s, n, r$) resembling those of JNW spacetime. Thus the best-fit values are listed in Tab. \ref{table2}.

\section{Conclusions\label{Sec:7}}

In this paper, we have studied the motion of a massive particle in the presence of scalar and gravitational fields, focusing on the Janis-Newman-Winicour (JNW) naked singularity solution. By deriving the equations of motion and the effective potential for a massive particle, we analyzed the innermost stable circular orbit (ISCO) radius in this spacetime. The ISCO radius, crucial for understanding dynamics around compact objects, was shown to be influenced by the scalar field and interaction parameter $g_s$. Our findings indicate that the ISCO radius decreases with positive $g_s$ and increases with negative $g_s$, highlighting the significant impact of scalar interactions on orbital dynamics. 

We have investigated the dynamics of massive particles in the presence of a scalar field including the effects of radiation reaction. The particles emit radiation due to their non-zero four-acceleration, with the intensity of the radiation depending on the coupling parameter $g_s$. By incorporating a radiation reaction term into the equation of motion, we derived a modified form that simplifies the analysis. Numerical calculations showed that the radiation reaction term has a minimal impact on the particle's trajectory. Our results indicate that the scalar field generates an attractive force, while the radiation reaction exerts a repulsive force. The trajectory of a massive particle around a naked singularity in the JNW spacetime is influenced by both these effects, as shown in our numerical simulations.

The oscillatory motion of test particles is a fundamental aspect of relativistic astrophysics, providing valuable information about the dynamics and characteristics of compact objects and their surrounding environments. We have examined the oscillatory motion of massive particles around compact gravitational objects within the JNW spacetime, influenced by the presence of a scalar field. Our analysis focused on radial and vertical oscillations, which are critical for understanding the dynamics and structure of spacetime around these objects. We derived the angular velocity and the expressions for specific energy and angular momentum of a massive particle in this spacetime, and further analyzed the effective potential governing the motion. We found that the presence of a scalar field significantly alters the fundamental frequencies of particle oscillations. Specifically, the radial and vertical frequencies were influenced by the scalar field's coupling parameter $g_s$ and the spacetime parameter $n$. Numerical results demonstrated that the interaction with the scalar field increases the fundamental frequencies, while higher values of the $n$ parameter lead to a decrease in these frequencies. Our findings provide insights into the complex behavior of particles near compact objects, emphasizing the importance of considering scalar fields in relativistic astrophysics. This work contributes to a deeper understanding of the oscillatory dynamics and potential observational signatures of particles in scalar field-influenced spacetimes.

We have also explored the oscillatory motion of test particles around compact gravitational objects, emphasizing its relevance to quasi-periodic oscillations (QPOs) observed in X-ray binaries. QPOs, believed to result from the oscillatory motion of matter in accretion disks around black holes or neutron stars, provide critical insights into the stability and structure of these disks. Understanding these oscillations aids in modeling emission spectra and variability, shedding light on the properties of black holes and neutron stars, including their mass, spin, and surrounding spacetime geometry. According to the RP model, twin-peaked QPOs are characterized by two main frequencies: the upper frequency ($\nu_U$) corresponding to the particle's orbital frequency, and the lower frequency ($\nu_L$) representing the difference between the orbital and radial oscillation frequencies. This model provides a framework for analyzing QPO phenomena in astrophysical systems with strong gravitational fields. Our analysis showed the radial dependence and relationships between the upper and lower frequencies of twin-peaked QPOs. The results indicate that increasing the parameters $n$ and $g_s$ shifts the frequency ratio of 3:2 QPOs closer to the naked singularity. Additionally, while increasing $n$ decreases both frequencies, increasing $g_s$ has the opposite effect. 

Finally, we analyzed four X-ray binary systems to constrain the parameters associated with the Janis-Newman-Winicour (JNW) solution by examining Quasi-Periodic Oscillations (QPOs) data from these systems. The celestial objects under study included XTE J1550-564, GRO J1655-40, and GRS J1915+105. Using Markov Chain Monte Carlo (MCMC) analysis with the Python library `emcee`, we derived constraints on the JNW parameters based on the relativistic precession (RP) model. Our analysis utilized Gaussian priors within specified boundaries for the parameters. The likelihood function, combining upper and lower frequency data, was used to derive the posterior distributions of these parameters. The best-fit values were obtained through this MCMC analysis, providing insights into the mass, coupling parameter $ g_s $, and the parameter $ n $ for each X-ray binary system. The obtained results demonstrate the constraints on the JNW parameters for each microquasar. It is also presented the best-fit parameter values derived from the QPOs data, reflecting the effectiveness of the RP model in describing the QPO phenomena in these X-ray binaries. This study enhances our understanding of the dynamics and properties of compact objects in strong gravitational fields.

\bibliography{Ref}
\end{document}